**Stresses, energy flow and energy density of gravitational nature**

A. LOINGER

Dipartimento di Fisica, Università di Milano

Via Celoria, 16 –  20133 Milano, Italy

**Summary.** – Two arguments which show the validity of the concept of gravitational energy put forward by Lorentz and by Levi-Civita.

**1**. – As is known, the characteristic conditions for the uniformity (i.e. homogeneity and isotropy) of a spacetime are the same of the existence conditions of an *integral* form of the conservation laws for the mass tensor $T_{jk}$, $(j,k= 0,1,2,3)$ [1]. A uniform spacetime is characterized by a Riemann curvature tensor of the following kind:

$$(1.1) \qquad R_{jkmn} = K(g_{km}g_{jn} - g_{jm}g_{kn}) \ ,$$

where $K$ is a constant. We are here interested in the case $K \neq 0$ [2]. The Einstein field equations are

$$(1.2) \qquad R_{jk} - \frac{1}{2} g_{ik} R = -\kappa T_{jk} \ ,$$

with

$$(1.2') \qquad T_{jk} = \frac{3K}{\kappa} g_{ik} \ ;$$

the spatio-temporal interval can be written in the simple form

$$(1.3) \qquad ds^2 = \frac{c^2 dt^2 - dx^2 - dy^2 - dz^2}{\left[1 - \frac{1}{4}K(c^2 t^2 - x^2 - y^2 - z^2)\right]^2} \ ,$$

which allows a transformation group with 10 parameters. The expression (1.3) represents our uniform spacetime on a pseudo-Euclidean manifold [3].

Evidently, the existence of an integral form of the conservation laws for the mass tensor (1.2') entails the existence of an integral form of similar conservation laws for the tensor $A_{jk} \equiv (1/\kappa)(R_{jk} - \frac{1}{2} g_{jk} R)$: we have thus *verified* in this particular, but interesting, case the validity of the interpretation of the gravitational tensor $A_{jk}$ which was proposed by Lorentz in 1916 [4] and, quite independently, by Levi-Civita in 1917 [5]. Indeed, according to these Authors the tensor $A_{jk}$ describes simply the stresses, the energy flow and the energy density of gravitational origin.

Levi-Civita emphasizes the strict analogy between the so-called d'Alembert principle of Newtonian dynamics, which says that the "lost" forces − i.e. the directly applied forces and the inertial ones − balance each other, and the relativistic assertion according to which the nature of the spacetime interval is always such to balance all mechanical actions: in fact, the sum of the mass tensor $T_{jk}$ and of the inertial-gravitational tensor $A_{jk}$ vanishes identically.

Einstein admitted that the above definition of Lorentz and Levi-Civita has an unquestionable *logical* soundness, but objected that it implies this *physically* peculiar consequence: if the total energy of a closed system must be always equal to zero, the conservation of this value does not assure the further existence of the system under any form whatever. Now, as a matter of fact, Einstein's objection was doomed to a total sterility, because the very structure of the relativistic formalism does not allow any intrinsic, *tensorial* definition of the gravitational energy different from that proposed by Lorentz and Levi-Civita, and the various *pseudo* (i.e. false) tensors put forward by several physicists are geometrically insignificant and physically unsatisfactory.

It is worth notice that in the last years Cooperstock, investigating some special questions, has attained to a conception of the gravitational energy which is very similar to Lorentz's and Levi-Civita's [6].

**1bis**. − Obviously, within any suitably limited region **L** of a given spacetime the Riemann curvature tensor can be approximated with a formula like (1.1). This means that there exists within **L** an integral form of the conservation laws both for $T_{jk}$ and for $A_{jk}$ : thus, the standpoint of Lorentz and Levi-Civita is expressively strengthened.

**2**. − As is known, it is a basic property of the Lagrangean formalism that if we perform the variation $\delta I$ of the action integral $I$ with respect to the components of the metric tensor (keeping the $g_{jk}$'s and their first derivatives constant on the spacetime boundary of the integration region $D$), we obtain the following result:

$$(2.1) \qquad \delta I = \int_D V^{jk} \sqrt{-g}\ \delta g_{jk}\ \mathrm{d}x,$$

where $V^{jk}$ is the *energy tensor* of the considered system. Now, eq. (2.1) holds even if our system is identical with any gravitational field $g_{jk}$ ; but in this case we have

$$(2.2) \qquad V^{jk} = A^{jk}.$$

**3**. − We have seen that for spacetimes of constant curvature the definition of gravitational energy due to Lorentz [4] and Levi-Civita [5] can be corroborated in a precise way. We have also seen that there exists a very general argument in favour of the interpretation of these Authors.

And now a grand finale. Levi-Civita concludes his memoir by observing that his definition of the gravitational energy **excludes** *a priori* the physical existence of a gravitational radiation and of other purely gravitational phenomena. In fact, by virtue of the field equations (a generalized d'Alembert principle)

(3.1) $$T_{jk} + A_{jk} = 0 \;,$$

when the mass tensor $T_{jk}$ vanishes, the same must happen to the *gravitational energy tensor* $A_{jk}$. "This fact entails a total lack of stresses, of energy flow and also of a simple energy localization" [5]. [7].

We have here the maximal generalization (*ante litteram*) of a remarkable Serini's theorem [8]: in a total absence of "matter" (not even "material" singularities be present), the spacetime must be **always** Minkowskian.

> "Fort! ins Land der Philister ihr Füchse,
> [mit brennenden Schwänzen,
> Und verderbet der Herrn reife papierene Saat".
> Goethe and Schiller